\documentclass{ws-procs10x7}
\usepackage{balance}

\newcolumntype{d}[1]{D{.}{.}{#1}}

\newcommand{\beq}[0]{\begin{equation}}
\newcommand{\eeq}[0]{\end{equation}}

\newcommand{\oa}{${\cal O}(\alpha)$}

\newcommand{\be}{\begin{equation}}
\newcommand{\ee}{\end{equation}}
\newcommand{\bea}{\begin{eqnarray}}
\newcommand{\eea}{\end{eqnarray}}
\newcommand{\myref}[1]{(\ref{#1})}
\newcommand{\BABAYAGA}{$\tt BABAYAGA$}
\newcommand{\BHWIDE}{$\tt BHWIDE$}
\newcommand{\LABSPV}{$\tt LABSPV$}
\makeindex
\begin{document}

\title{Matching Parton Shower and matrix elements in QED}

\author{G. Balossini}
\address{Dipartimento di Fisica Nucleare e Teorica, v. A. Bassi, I-27100 Pavia, Italy\\
E-mail: Giovanni.Balossini@pv.infn.it}

\author{C.M.~Carloni Calame$^*$, O.~Nicrosini$^{\diamond}$ \and 
F.~Piccinini$^{\dagger}$}
\address{INFN Sezione di Pavia, v. A. Bassi, I-27100 Pavia, Italy\\
$^*$E-mail: Carlo.Carloni.Calame@pv.infn.it\\
$^{\diamond}$E-mail: Oreste.Nicrosini@pv.infn.it\\
$^{\dagger}$E-mail: Fulvio.Piccinini@pv.infn.it
}

\author{G.~Montagna}
\address{Dipartimento di Fisica Nucleare e Teorica and INFN Sezione di Pavia, 
v. A. Bassi, I-27100 Pavia, Italy\\
E-mail: Guido.Montagna@pv.infn.it}


\twocolumn[\maketitle\abstract{We report on a high-precision calculation of the Bhabha process in QED, of interest for precise luminosity
determination of low-energy electron-positron colliders. 
The calculation is based on the matching of exact next-to-leading 
order corrections with a Parton Shower algorithm. 
The structure of the algorithm (implemented in an improved version 
of the event generator \BABAYAGA) is illustrated, with a discussion on 
the resulting theoretical uncertainty, of the order of 0.1\%.}

\keywords{QED; Bhabha; luminosity; next-to-leading corrections; Parton Shower.}
]

\section{Introduction}
\label{intro}
The measurement of the ratio 
$R = \sigma(e^+  e^- \to {\rm hadrons})/\sigma(e^+ e^- \to \mu^+ \mu^-)$ 
at flavour factories, such as DA$\Phi$NE, VEPP-2M, BES, KEK-B, 
PEP-II and CLEO, 
is of primary importance for the precise determination 
of the anomalous magnetic moment of the muon, $(g-2)_{\mu}$, and 
of the running of the electromagnetic coupling $\alpha_{QED}(Q^2)$. The 
cross section values entering $R$ are affected 
by the uncertainty on the knowledge of the machine luminosity, 
which is, in turn, related to the uncertainty on the theoretical knowledge 
of the cross section of a reference QED process, 
typically large angle Bhabha scattering, $\mu^+ \mu^-$ and 
$\gamma \gamma$ production. 
To keep under control such an uncertainty, high-precision calculations 
of these QED processes and relative Monte 
Carlo generators, are required. 
Large-angle Bhabha scattering is
of particular interest because of its large cross section and
its clean experimental signature. To 
simulate the experimentally relevant distributions and calculate the
cross section of the Bhabha process, KLOE and CLEO 
collaborations make use of the QED Parton Shower (PS) generator
\BABAYAGA, developed in Refs.~\cite{babayaga,ips} with a 
precision of 0.5\%. At present a reduction of such a 
theoretical systematics is needed for several reasons. 
First, the experimental luminosity error quoted by KLOE 
is presently 0.3\% \cite{kloeupdated}. 
Secondly, the measurement
of the hadronic cross section in the $\pi^+ \pi^-$ channel at VEPP-2M
has achieved a total systematic error of 0.6$-$1\% in the region of
the $\rho$ resonance~\cite{cmd2snd}, which requires,
in turn, an assessment of the collider luminosity at the level of
0.1\%. Last but not least, Charm and $B$-factories will reach in 
the near future the error of 1\% in the luminosity measurement.

At the 0.1\% level the non logarithmic contributions present in exact 
next-to-leading (NLO) perturbative calculations as well as the resummed 
leading logarithmic contributions taken into account in the PS  
approach are expected to be relevant. A matching algorithm which allows 
to incorporate the next-to-leading terms within the PS structure 
of an event generator such as {\tt BABAYAGA}, without double counting 
at first order in $\alpha$
of the leading corrections already accounted for by the PS, has  
been developed in Ref.~\cite{babayaga@nlo} and will be reviewed in the 
following. An estimate of the remaining theoretical uncertainty is 
also discussed.

\section{Matching NLO corrections with Parton~Shower} 
\label{matching}

A general expression for the cross section with the emission of an
arbitrary number of photons, in leading-log (LL) approximation,
can be cast in the following form:
\be
d\sigma^{\infty}_{LL}=
{\Pi}(Q^2,\varepsilon)~
\sum_{n=0}^\infty \frac{1}{n!}~|{\cal M}_{n,LL}|^2~d\Phi_n
\label{generalLL}
\ee
where ${\Pi}(Q^2,\varepsilon)$ is the Sudakov form-factor accounting for the
soft-photon (up to an energy equal to $\varepsilon$ in units of the
incoming fermion energy $E$) and virtual emissions, $\varepsilon$ is
an infrared separator 
dividing soft and hard radiation and $Q^2$ is
related to the energy scale of the process.
$|{\cal M}_{n,LL}|^2$ is the squared amplitude in LL
approximation describing the process with the emission of $n$ hard
photons, with energy larger than $\varepsilon$ in units of $E$.
$d\Phi_n$ is the exact 
phase-space element of the process (divided by the incoming flux
factor), with the emission of $n$ additional photons
with respect to the Born-like final-state configuration. 

According to the factorization theorems of soft and/or collinear
singularities,
the squared amplitudes in LL approximation can be written in a factorized
form. In the following, for the sake of clarity and without loss of
generality, we write photon emission formulas as if only one external
fermion radiates, being the generalization to the real case 
straightforward when including the suited combinatorial factors. 
The one-photon emission squared amplitude in LL approximation can be
written in factorized form as
\be
|{\cal M}_{1,LL}|^2=\frac{\alpha}{2\pi}\frac{1+z^2}{1-z}I(k)
~|{\cal M}_0|^2~
\frac{8\pi^2}{E^2 z (1-z)}
\label{onegammaLL}
\ee
where $1-z$ is the fraction of the fermion energy $E$ carried by the
photon, $k$ is the photon four-momentum,
$I(k)$ is a function describing the angular spectrum of the photon. 

The cross section calculated in Eq.~\myref{generalLL} has the
advantage that the photonic corrections, in LL approximation, are
resummed up to all orders of perturbation theory. On the other side,
the weak point of the formula~\myref{generalLL} is that its
expansion at \oa\ does not coincide with an exact \oa\ (NLO) result, being
its LL approximation. In fact
\bea
d\sigma^{\alpha}_{LL} &\equiv&
\left[
1+C_{\alpha,LL}
\right] |{\cal M}_0|^2 d\Phi_0
+
|{\cal M}_{1,LL} |^2 d\Phi_1 \nonumber \\
&&
\label{LL1}
\eea
whereas an exact NLO cross section can be always cast in the form
\be
d\sigma^{\alpha}
=
\left[
1+C_{\alpha}
\right] |{\cal M}_0|^2 d\Phi_0
+
|{\cal M}_{1} |^2 d\Phi_1
\label{exact1}
\ee
The coefficient $C_{\alpha}$ contains the complete virtual
\oa\ and the \oa\ soft-bremsstrahlung squared matrix elements, in units of
the Born squared amplitude,
and $|{\cal M}_1|^2$ is the exact squared matrix element with the
emission of one hard photon.
We remark that $C_{\alpha,LL}$ has the same logarithmic structure as
$C_\alpha$ and that $|{\cal M}_{1,LL}|^2$ has the same singular
behaviour of $|{\cal M}_1|^2$.

In order to match the LL and NLO calculations, we introduce
the correction factors, which are by construction
infrared safe and free of collinear logarithms,
\begin{eqnarray}
F_{SV}~&=&~
1+\left(C_\alpha-C_{\alpha,LL}\right), \nonumber \\
F_H~&=&~
1+\frac{|{\cal M}_1|^2-|{\cal M}_{1,LL}|^2}{|{\cal M}_{1,LL}|^2}
\label{FSVH}
\end{eqnarray}
and we notice that the exact \oa\ cross section can be expressed, up
to terms of ${\cal O}(\alpha^2)$,
in terms of its LL approximation as
\begin{eqnarray}
d\sigma_\alpha~&=&~
F_{SV} (1+C_{\alpha,LL} ) |{\cal M}_0|^2 d\Phi_0 \nonumber \\
~&+&~
F_H |{\cal M}_{1,LL}|^2 d\Phi_1
\label{matchedalpha}
\end{eqnarray}
Driven by Eq.~\myref{matchedalpha}, Eq.~\myref{generalLL} can be improved
by writing the resummed cross section as
\begin{eqnarray}
&d\sigma&^{\infty}_{matched}=
F_{SV}~\Pi(Q^2,\varepsilon) \nonumber \\
&\times&\sum_{n=0}^\infty \frac{1}{n!}~
\left( \prod_{i=0}^n F_{H,i}\right)~
|{\cal M}_{n,LL}|^2~
d\Phi_n
\label{matchedinfty}
\end{eqnarray}
The expansion at \oa\ of Eq.~\myref{matchedinfty} coincides now with
the exact NLO cross section Eq.~\myref{exact1} and
all higher order LL contributions
are the same as in Eq.~\myref{generalLL}.

Eq.~\myref{matchedinfty} is our master formula for the matching between
the exact \oa\ calculation and the QED resummed PS
cross section, according to which we also generate events.
The correction factors of Eq.~\myref{FSVH} can in principle make the
differential cross section of Eq.~\myref{matchedinfty} negative in
some point, namely where the PS approximation is less accurate. 
Nevertheless, we verified
that this never happens when considering typical event selection criteria for
luminosity at flavour factories.

\subsection{Vacuum polarization}
\label{vpsubsection}
Besides the photonic radiative corrections 
considered above, also the vacuum
polarization effects must be included in the master 
formula~\myref{matchedinfty}, in order to reach the required 
theoretical accuracy. They are accounted for by replacing
the fine structure constant $\alpha\equiv\alpha(0)$ with 
$\alpha(q^2)=\alpha/(1-\Delta\alpha(q^2))$, according to the algorithm 
described in Ref.~\cite{babayaga@nlo}. $\Delta\alpha(q^2)$
is the fermionic contribution to the photon self-energy: the leptonic
and top-quark one-loop contributions can be calculated analytically in
perturbation theory, while the remaining five quarks (hadronic)
contribution, $\Delta\alpha^{(5)}_{hadr}$, has to be extracted from
data. To evaluate $\Delta\alpha^{(5)}_{hadr}$ we use the $\tt HADR5N$
routine~\cite{jeg-2003,hadr5}.

Going beyond the Born-like approximation, the cross section corrected
at \oa\ including also vacuum polarization can be written as
$\sigma^\alpha_{VP}=\sigma_{0,VP} + \sigma^\alpha_{SV} +
\sigma^\alpha_{H}$, where $\sigma^\alpha_{SV}$ and $\sigma^\alpha_{H}$
are the soft plus virtual and the hard photon \oa\ corrections of
photonic origin. We can go a step further and include vacuum
polarization in those terms, in order to include also part of the
${\cal O}(\alpha^2)$ factorizable corrections.  

Furthermore, we add to the Born amplitude also the $Z$ exchange diagrams: their
effect is really tiny and negligible at low energies, but can become
more important (up to 0.1\%) around 10 GeV. 

\section{Theoretical uncertainty}
\label{theoacc}
Since different implementations of radiative corrections beyond exact 
${\cal O}(\alpha)$ contributions differ by higher order effects, 
a hint of the missing radiative corrections which dominate 
the theoretical accuracy can be given by comparing the predictions 
of \BABAYAGA\ with independent event generators, such as 
\BHWIDE\cite{bhwide}, \LABSPV\cite{labspv} and {\tt MCGPJ}\cite{mcgpj}. 
As shown in Ref.~\cite{babayaga@nlo}, the results of these comparisons 
are very satisfactory, with differences between \BABAYAGA\ and \BHWIDE\  below 
0.1\% on cross sections and ranging up to 1\% only in the tails of 
some distributions where the statistics is low. 

A firmer estimate of the theoretical accuracy could be given by comparison 
with a complete two-loop calculation, which is however not available yet. 
However, recently there has been important progress towards the full  
${\cal O}(\alpha^2)$ calculation. At present, two different partial 
contributions have been calculated: the complete virtual two-loop photonic 
corrections (in the limit $Q^2 \gg m_e^2$, with $Q^2 = s$, $t$, $u$) plus real 
radiation in soft approximation~\cite{penin} and the virtual $N_f=1$ fermionic 
contribution inclusive of finite mass terms~\cite{boncianietal}. 
In Ref.~\cite{babayaga@nlo} the 
terms in \BABAYAGA\  corresponding to these two approximations of the 
complete ${\cal O}(\alpha^2)$ calculation have been extracted and compared, 
showing excellent agreement. The relative differences 
don't exceed the 0.03\% level. A careful inspection of the analytical 
expressions of the differences shows that all logarithmic terms 
of infrared origin present in \BABAYAGA\  have the same coefficients as in the 
two-loop perturbative calculations, 
with the exception of small terms suppressed by powers of $m_e^2/s$. 

Other ${\cal O}(\alpha^2)$ contributions not considered in \BABAYAGA\ 
(and therefore sources of theoretical uncertainty) 
are the light pair corrections and the soft plus virtual 
${\cal O}(\alpha)$ corrections to the real hard radiation. The impact 
of the former contribution has been estimated for a sample of typical 
event selection and energies, with $t$-channel virtual 
LL approximated formulae\cite{virtualpairs} 
and real pair emission in soft approximation\cite{realsoftpairs}. 
The impact of such 
corrections has been found below the 0.05\% level. 
Exact perturbative results for the soft plus virtual 
corrections to real hard radiation are not available yet for the complete 
$s+t$ Bhabha process, which is of interest for the case of low energy 
$e^+ e^-$ colliders. They have been calculated separately for the $t$- and 
$s$-channels and their impact studied at LEP conditions. 
Taking into account of such 
experience and that \BABAYAGA\  contains all the infrared enhanced terms, 
the size of these ${\cal O}(\alpha^2)$ corrections 
can be estimated to be smaller than 0.05\%. 
Adding in quadrature the perturbative sources of theoretical uncertainty 
would give a theoretical error of the order of 0.1\% for all considered 
event selections and energies, from 1 to 10~GeV. However this value 
underestimates the total error associated with \BABAYAGA\  
at Charm and $B$-factories because the routine 
{\tt HADR5N}, which gives the vacuum polarisation corrections, 
produces large errors around the $J/\psi$ and $\Upsilon$ resonances, 
increasing the total theoretical error of \BABAYAGA\  at the 0.2\% level 
in those energy ranges.

{\emph{F.P. would like to thank the conveners 
of the electroweak parallel session 
for their kind invitation.}}

\end{document}